# High $q$-resolution neutron scattering technique using triple-axis spectrometers

GUANGYONG XU,[a]* P. M. GEHRING,[b] V. J. GHOSH[a] AND G. SHIRANE[a]

[a]*Physics Department, Brookhaven National Laboratory, Upton, NY 11973, and [b]NCNR,*

*National Institute of Standards and Technology, Gaithersburg, Maryland, 20899. E-mail:*

*gxu@bnl.gov*

## Abstract

We present a new technique which brings a substantial increase of the wave-vector $q$-resolution of triple-axis-spectrometers by matching the measurement wave-vector $q$ to the reflection $\tau_a$ of a perfect crystal analyzer. A relative Bragg width of $\delta/Q \sim 10^{-4}$ can be achieved with reasonable collimation settings. This technique is very useful in measuring small structural changes and line broadenings that can not be accurately measured with conventional set-ups, while still keeping all the strengths of a triple-axis-spectrometer.

## 1. Introduction

Triple-axis-spectrometers (TAS) are widely used in both elastic and inelastic neutron scattering measurements to study the structures and dynamics in condensed matter. It has the flexibility to allow one to probe nearly all coordinates in energy ($\hbar\omega$) and momentum ($q$) space in a controlled manner, and the data can be easily interpreted (Bacon, 1975; Shirane *et al.*, 2002).

The resolution of a triple-axis-spectrometer is determined by many factors, including the incident ($E_i$) and final ($E_f$) neutron energies, the wave-vector transfer $Q$, the monochromator and analyzer mosaic, and the beam collimations, etc. This has been studied in detail by Cooper & Nathans (1967), Werner & Pynn (1971) and Chesser & Axe (1973). It has long been known that when the measurement wave-vector is close to that of the monochromator ($\tau_m = 2\pi/d_m$), good $q$-resolution can be achieved. This is called the "focusing condition". However, since there is very little freedom to vary the monochromator d-spacing in an experiment, it is usually not possible to achieve focusing near the wave-vector of interest.

Recently experiments and calculations have shown that by matching the reflection of the analyzer ($\tau_a = 2\pi/d_a$) with the measured wave-vector, a similar focusing condition can be achieved. The improvement in $q$-resolution is particularly great when the analyzer is a perfect crystal with very fine mosaic. However this is not the case in a conventional triple-axis-spectrometer where the monochromator and analyzer crystals are deliberately distorted so that their mosaic (typically $\sim$30' to 1°) matches the beam collimations (typically 10' to 100'). In most cases using a perfect crystal as analyzer/monochromator only results in much lower intensity because of the large primary extinction, and will not improve the resolution significantly. This is because the much coarser beam collimations control the resolution. In case of elastic scattering measurements, however, the intensity is generally not a major concern. The Bragg intensity from a reasonable size single crystal ($\sim$ a few grams) can easily reach $10^4$ counts per second. It is therefore feasible to trade off intensity for higher instrumental resolution, if the analyzer side focusing condition can be satisfied.

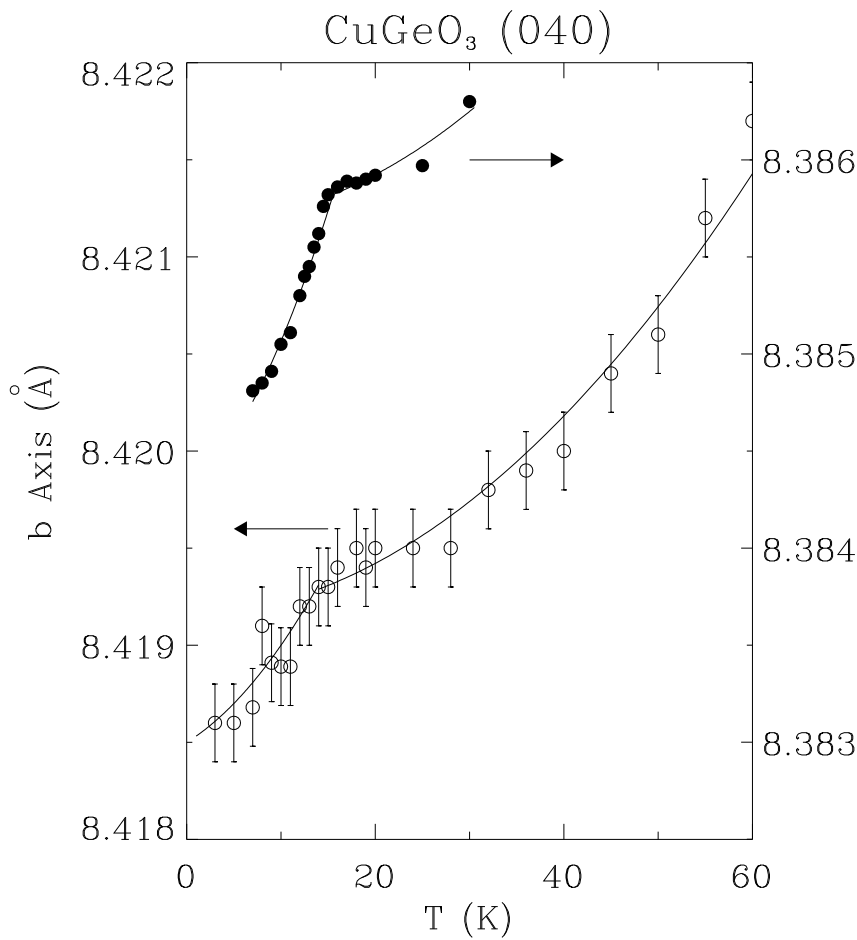

Fig. 1. Temperature dependence of the $b$ lattice constant of CuGeO$_3$ measured at $\mathbf{Q} = (0, 4, 0)$, using high $q$-resolution neutron scattering (Ge(220) analyzer, $E_i = 7.48$ meV, 10'-10'-10'-10' collimations, open circles (Lorenzo $et\ al.$, 1994)), and x-ray diffraction (solid circles (Harris $et\ al.$, 1994)).

Lorenzo $et\ al.$ (1994) used this new technique to measure the lattice parameter of CuGeO$_3$. The reciprocal lattice vector associated with the (040) Bragg peak of CuGeO$_3$ matches nearly perfectly with Ge (220) reflection ($\tau_{Ge(220)} = 3.1414$ Å$^{-1}$). The instrument set-up employed a PG(002) monochromator tuned to an incident neutron energy $E_i = 7.48$ meV, a perfect crystal Ge (220) analyzer, and beam collimations of 10'-10'-10'-10'. A $q$-resolution of 0.002 Å$^{-1}$ was achieved with this set-up. Fig. 1 compares the lattice parameter of CuGeO$_3$ measured by Lorenzo $et\ al.$ (1994) with that of high resolution x-ray diffraction measurements of Harris *et al.* (1994). The temperature dependences of these results agree perfectly (lattice spacings offset by 0.04 Å), demonstrating the potential of this new technique in measuring lattice parameters to a relative accuracy of $\sim 10^{-4}$, which is approaching that of high resolution x-ray diffraction measurements.

This use of the (220) reflection of a perfect Ge crystal analyzer has been further exploited by Ohwada *et al.* (2001) and Gehring *et al.* (2003) in measuring the structure of $PbXO_3$ type relaxor ferroelectrics. These are perovskites with lattice parameter $a \approx 4.0$ Å. The length of the reciprocal lattice vector $q_{200} \approx 3.14$ Å$^{-1}$ in these systems is very close to $\tau_{Ge(220)}$. The longitudinal Bragg full-width at half maximum (FWHM) in these measurements is about 0.003 Å$^{-1}$. Many other ferroelectric perovskite systems, such as $SrTiO_3$, have very similar lattice parameters and thus can be studied with this technique.

One example is shown in Fig. 2, where longitudinal scans along the (200) reflection of single crystal $SrTiO_3$ are plotted. The data were taken on the BT9 triple-axis-spectrometer located at the NIST Center of Neutron Research (NCNR). The monochromator is a PG(002) single crystal with a mosaic of $\sim 35$' in both horizontal and vertical directions. The incident neutron energy was $E_i = 14.7$ meV. The open circles represent data taken using a perfect Ge(220) crystal as analyzer, and reasonable beam collimations of 10'-40'-20'-40'. The Bragg peak is extremely sharp, with FWHM $\delta q/Q \approx 6 \times 10^{-4}$, almost one order of magnitude better than that obtained using a PG(002) analyzer of mosaic $\sim 35$', and the finest collimations available (10'-10'-10'-10', solid circles in Fig. 2).

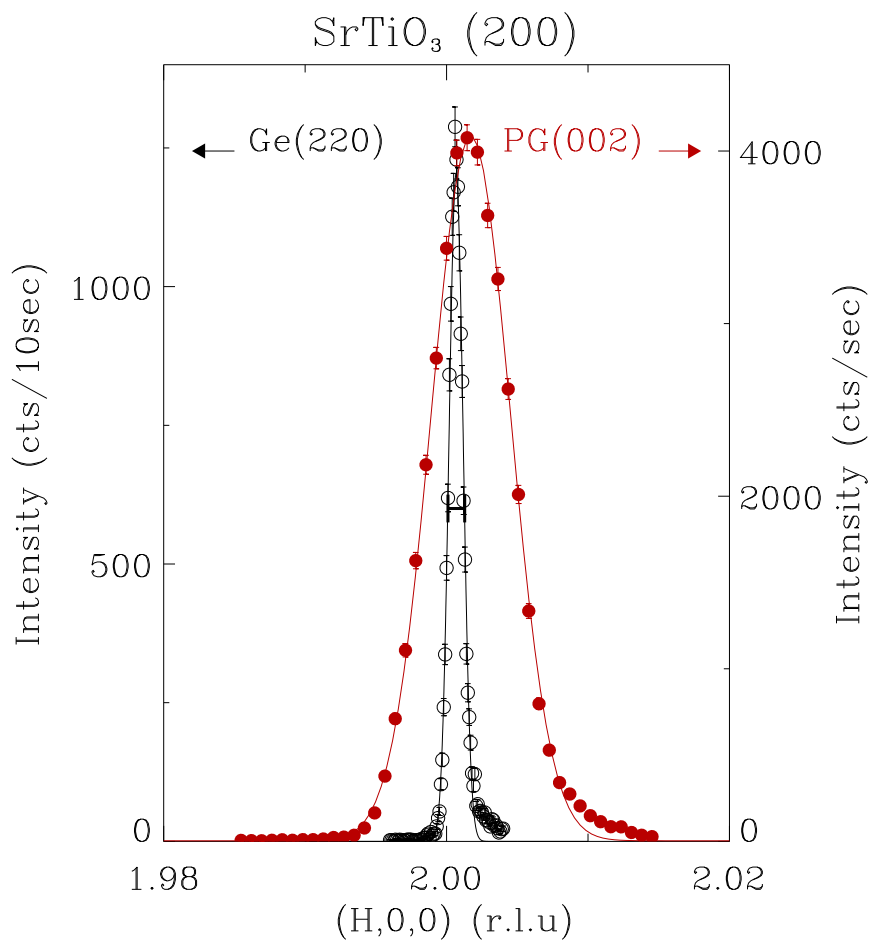

Fig. 2. A single crystal SrTiO$_3$ longitudinal scan measured at (200) with a PG(002) analyzer, 10'-10'-10'-10' collimations, (solid circles) and a Ge(220) perfect crystal analyzer, 10'-40'-20'-40' collimations (open circles). The resolution for the Ge(220) set-up is represented by the horizontal bar.

In this paper we describe this new technique in detail using a perfect Ge crystal (mosaic $\leq 1'$) as analyzer. When the analyzer side focusing condition is satisfied, and with a proper choice of collimations, a more than one order of magnitude improvement can be obtained in the longitudinal Bragg width (compared to PG(002) (mosaic $\sim$35') analyzer) with reasonable intensities. By using different reflections of perfect crystal analyzer, different Bragg peaks of the sample crystal can be studied. With this technique, we can achieve a reasonably good $q$-resolution, while retaining the strength and flexibility of a triple-axis-spectrometer.

## 2. Resolution Calculations

When measuring lattice distortions and structural phase transitions, the magnitude of the Bragg peak splitting, peak width broadening, and other related effects are usually proportional to the length of the wave vector. Therefore, the relative Bragg width $\delta q/Q$ is of more importance than the absolute Bragg width $\delta q$ itself. In this section, the relative Bragg width $\delta q/Q$ is calculated for different instrument set-ups and effective sample mosaics. The calculations are based on the formulas derived by Cooper & Nathans (1967), Werner & Pynn (1971) and Chesser & Axe (1973). The typical instrument set-up of the BT9 triple-axis-spectrometer at NCNR was chosen for the calculation/simulations. The monochromator is a vertically focusing PG(002) crystal with mosaic of ~35'.

With the new technique, a sharp longitudinal Bragg resolution can only be achieved around a small range of $Q$ that matches the perfect crystal analyzer reciprocal lattice vector $\tau_a = 2\pi/d_a$. It is much easier to satisfy the focusing condition on the analyzer side than on the monochromator side because the analyzer of a conventional TAS can be easily changed.

We will first focus our discussion on one specific perfect crystal analyzer set-up, the Ge(220) reflection, which can be directly compared to the PG(002) analyzer, mostly because their focusing conditions are similar. A detailed comparison is given in Fig. 3. Here we have used beam collimations of 10'-40'-20'-40' and an incident energy of 14.7 meV for the Ge(220) set-up. For the PG(002) set-up, we used the best available beam collimation 10'-10'-10'-10' in the calculations. If the sample itself is a perfect single crystal (mosaic $\leq 1'$), the relative longitudinal Bragg width using a PG(002) analyzer is about $\delta q/Q \sim 10^{-3}$ to $10^{-2}$ in the range of 1 Å$^{-1} < Q < 6$ Å$^{-1}$, with a minimum of $\sim 2.7 \times 10^{-3}$. By switching to the perfect Ge crystal (220) ($\tau_{Ge(220)} = 3.14131$ Å$^{-1}$) reflection as the analyzer, the longitudinal Bragg width near $Q = 3.14$ Å$^{-1}$ is improved by one order of magnitude, $\delta q/Q \approx 5.5 \times 10^{-4}$. This value is approaching the Bragg resolution of high energy x-ray scattering measurements, and about an order of magnitude better than that of a regular neutron powder diffractometer.

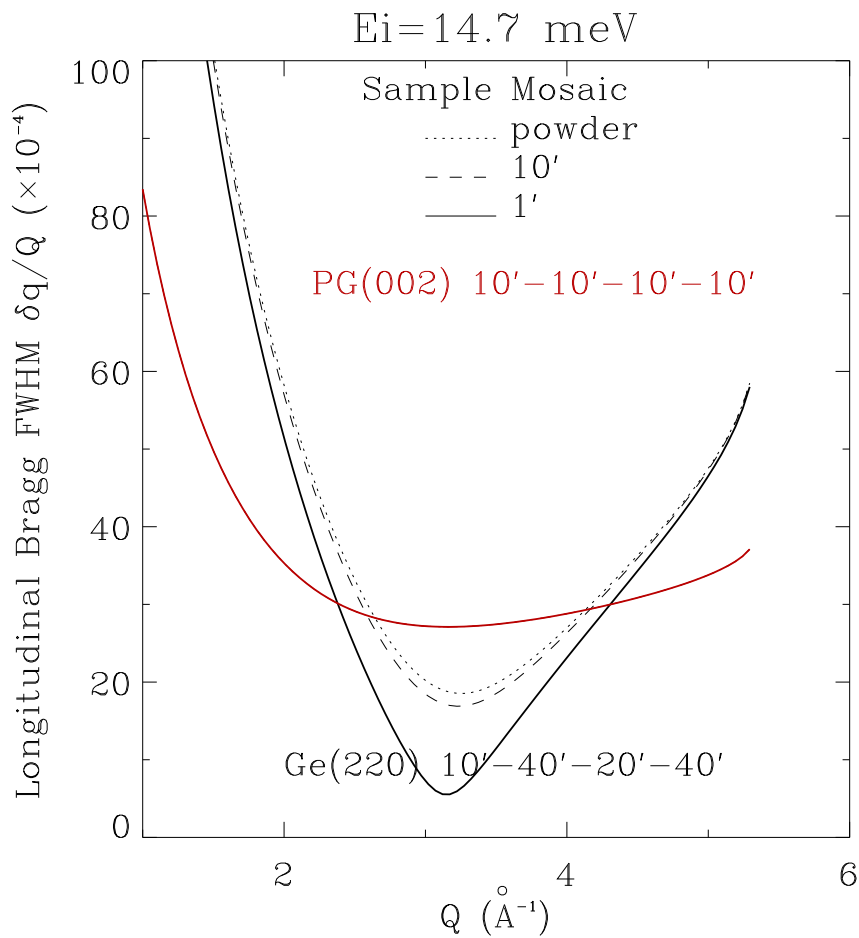

Fig. 3. Longitudinal Bragg width using a conventional PG(002) analyzer (red lines, beam collimations 10'-10'-10'-10') and a Ge(220) perfect crystal analyzer (black lines, beam collimations 10'-40'-20'-40'). The dashed and dotted lines are calculations assuming a 10' sample mosaic and a powder sample.

Now we consider the effect of non-zero sample mosaic on the Bragg width. The Bragg widths calculated for an imperfect single crystal sample (mosaic = 10') and a powder sample are plotted in Fig. 3 using dashed and dotted lines. The calculations show that the longitudinal Bragg width of the measurement is greatly affected by the sample mosaic, but only around those $q$ values close to $\tau_a$. The longitudinal width calculated for a sample crystal with 10' mosaic is $\delta q/Q \sim 1.7 \times 10^{-3}$, more than three times larger than that of a perfect single crystal sample. A better illustration of the effect of sample mosaic on the longitudinal resolution is shown in Fig. 4. Here we plot the elastic resolution ellipses around a PbXO$_3$ (200) Bragg peak ($q \approx 3.14$ Å$^{-1}$). For the Ge(220) setup, when sample mosaic increases, not only the width, but also the shape and orientation of the resolution ellipse changes. This results in a significant increase in the longitudinal Bragg width. Although it is still much better than that of a normal triple-axis set-up, this has to be taken into consideration during the measurements.

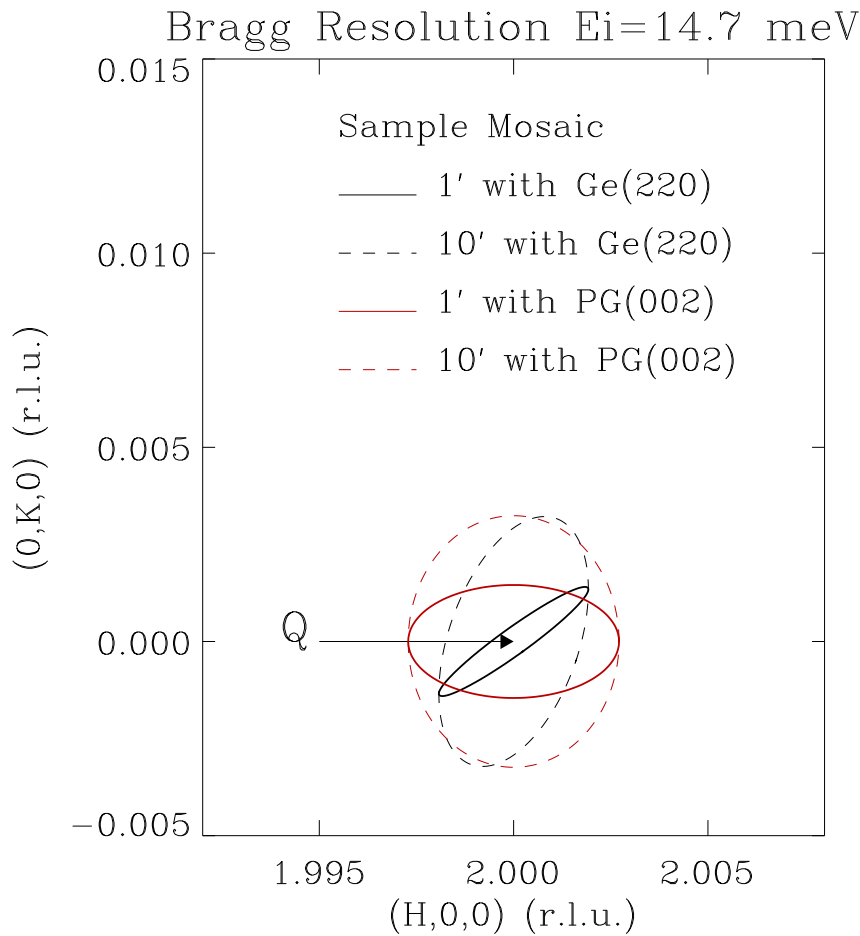

Fig. 4. Bragg resolution ellipse for PG(002) and Ge(220) analyzer set-ups. The beam collimations are 10'-10'-10'-10' for the PG set-up and 10'-40'-20'-40' for the Ge set-up. The dashed lines are results calculated assuming a sample mosaic $\eta = 10'$.

If the system being measured undergoes a structural phase transition that causes the crystal effective mosaic to increase, it will affect measurements in the longitudinal Bragg width as well. In this case, both longitudinal and transverse scans around the Bragg width of interest should be analyzed carefully, with resolution de-convoluted in the analysis in order to obtain the true mosaic/strain information. This is contrary to the case of most synchrotron high energy x-ray diffraction measurements. In the case of x-ray measurements, the resolution is more defined by the very fine mosaic of the monochromator and analyzer crystals (usually high quality Si crystals with mosaic on the order of $10^{-3}$ to $10^{-2}$ minutes), and the longitudinal Bragg width is unaffected by sample mosaic.

For wave vectors close to $q = \tau_a$, not only longitudinal, but also transverse Bragg widths of the measurements are greatly improved. Fig. 5 shows the transverse Bragg width of set-ups using PG(002) and perfect crystal Ge(220) analyzers. At $q \approx \tau_a$, the relative transverse resolution for the set-up with perfect crystal Ge(220) analyzer is only limited by the sample mosaic, yet that of the PG(002) set-up is much broader, $\delta q_\perp / Q \sim 1.5 \times 10^{-3}$ ($\sim 5'$). Therefore, this new technique with perfect Ge crystal analyzer is effective in not only detecting small strains (difference in the length of the Bragg wave vectors), but also small mosaic distortions and twinnings.

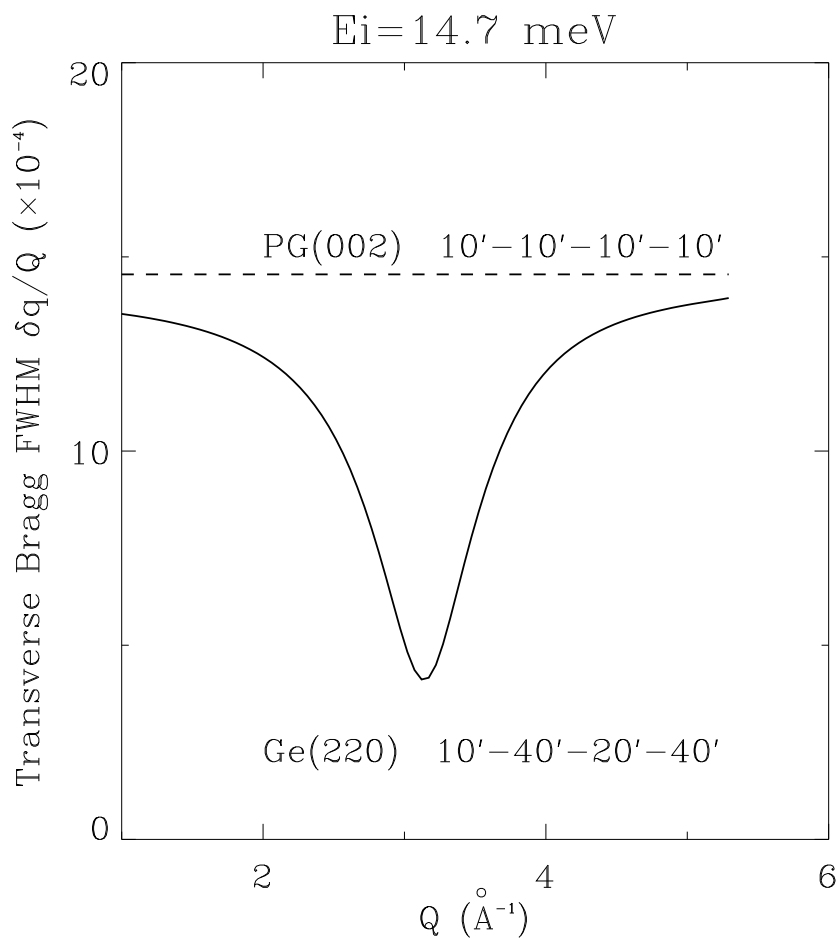

Fig. 5. Transverse Bragg resolution for PG(002) and Ge(220) analyzer set-ups. The collimations are 10'-10'-10'-10' for the PG set-up and 10'-40'-20'-40' for the Ge set-up. The sample mosaic is chosen to be $\leq 1'$.

In addition to sample mosaic, another contributing factor to the Bragg resolution are the beam collimations. In Fig. 6, longitudinal Bragg widths using a Ge(220) analyzer, with different beam collimations are shown. For $q$ close to $\tau_a$, the width is not affected at all by the beam collimations. This is a natural result since the Bragg width at this point is largely defined by the fine mosaic of the perfect Ge(220) analyzer crystal. When $q$ moves away from this optimum value, one starts to see the effect of resolution broadening by coarser beam collimations. When $q$ is 10% larger than $\tau_a$, the Bragg width can change by a factor of two, if the beam collimations change from 10'-10'-10'-10' to 10'-40'-20'-40. In the course of an experiment, if the wave vector to be measured is very close to $\tau_a$, a coarser collimation will help to increase the intensity while not sacrificing too much in resolution. On the other hand, if better $q$ resolution is essential over a range of $q$, then better collimations should be used to obtain better resolution when $q$ deviates slightly from the focusing condition.

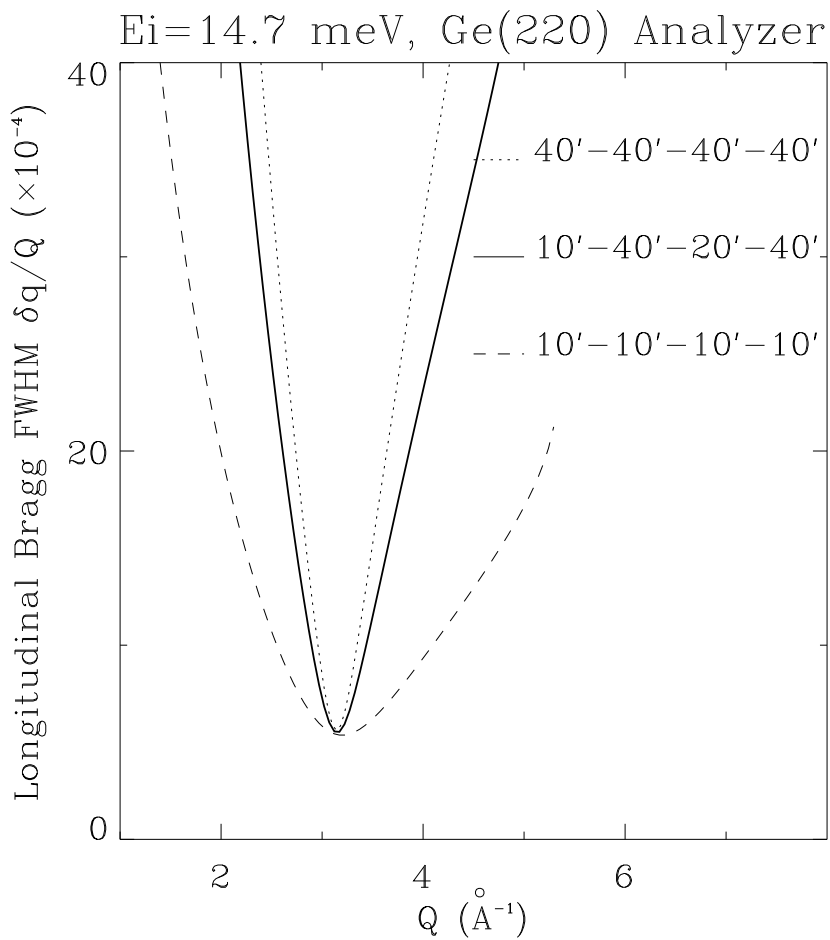

[ht]

Fig. 6. Longitudinal Bragg width using a Ge(220) perfect crystal analyzer, with different beam collimations.

More recently, we noticed that by using an analyzer reflection with an even larger $\tau$ value, a better relative Bragg width can be achieved. In Fig. 7, we show the relative longitudinal Bragg widths using different Ge reflections as analyzer crystal, with the same beam collimations 10'-40'-20'-40'. Here we see that that with different analyzer reflections, very good longitudinal Bragg widths ($\delta q/Q \sim 10^{-4}$) can always be achieved when the analyzer side focusing condition is satisfied. The relative Bragg width improves by almost a factor of two from the set-up using the Ge (220) reflection at $q \approx 3.14$ Å$^{-1}$, $\delta q/Q \approx 5.5 \times 10^{-4}$, to the Ge (004) reflection at $q \approx 4.44$ Å$^{-1}$, $\delta q/Q \approx 2.6 \times 10^{-4}$. More experimental examples will be shown in Section 4. When changing to reflections with even longer $\tau_{Ge}$ values, $\delta q/Q$ still improves, though not as much.

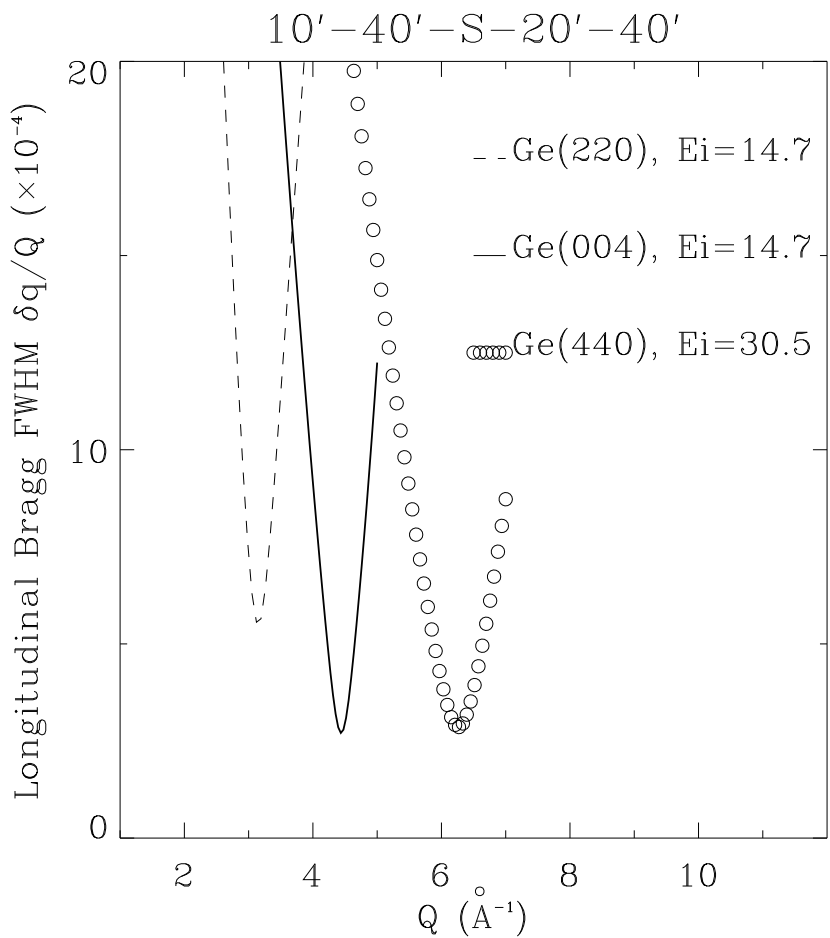

[ht]

Fig. 7. Longitudinal Bragg width with different analyzers and/or different incident neutron energies $E_i$.

One can always use other perfect single crystals, e.g., perfect Si crystals, perfect Cu crystals, etc., as analyzers, to adjust where the minimum of the Bragg width lies in $q$, depending on the needs of the experiment. Another choice of perfect crystal analyzers is SrTiO$_3$ (see Table 1 for different perfect crystal analyzers). The advantage of SrTiO$_3$ is that it has a per-ovskite structure and all its reflections are available as analyzer reflections, which provide much more choices of $\tau$. A second perfect crystal of the same compound being studied can also be used as the analyzer. Such a choice makes it possible to satisfy the focusing condition at every Bragg reflection of interest.

Table 1. *Reciprocal spacings $\tau = 2\pi/d$ for different perfect crystal analyzers. (PG (002) and (004) are also listed for comparison.)*

| Crystal Reflection | $\tau = 2\pi/d$ ( Å$^{-1}$) | SrTiO$_3$ Reflection | $\tau$( Å$^{-1}$) |
|---|---|---|---|
| PG(002) | 1.873 | (100) | 1.609 |
| Ge(111) | 1.924 | | |
| Si(111) | 2.004 | (110) | 2.276 |
| Cu(111) | 3.010 | (111) | 2.787 |
| Ge(220) | 3.141 | | |
| Si(220) | 3.272 | (200) | 3.218 |
| Cu(002) | 3.476 | | |
| Ge(311) | 3.683 | (210) | 3.560 |
| PG(004) | 3.747 | | |
| Si(311) | 3.837 | (211) | 3.941 |
| Ge(004) | 4.442 | (220) | 4.551 |
| Si(004) | 4.628 | | |
| Ge(331) | 4.841 | (300) | 4.827 |
| Cu(220) | 4.916 | | |
| Si(331) | 5.043 | (310) | 5.088 |

## 3. Intensity Simulations

Table. 2 shows results from simulations using the MCSTAS Monte Carlo code. We have performed simulations for different collimations using the Ge (220) and (004) reflections as analyzers. The relative intensities are calculated at the $q$ values satisfying the focusing condition, i.e., $q = 4.44$ Å$^{-1}$ for the Ge (004) set-up and $q = 3.14$ Å$^{-1}$ for the Ge (220) set-up.

Table 2. *Monte Carlo simulation results on relative intensities of various set-ups with different analyzer reflections and beam collimations. The incident energy is tuned to 14.7 meV. The monochromator is a PG(002) crystal with mosaic of $\sim 35'$. The intensities are calculated assuming sample mosaics of 1' and 60'.*

| Analyzer | Collimations | Relative intensity | | | |
|---|---|---|---|---|---|
| | | sample mosaic $\eta = 60'$ | | sample mosaic $\eta = 1'$ | |
| | | $Q = 3.14\,(\text{Å}^{-1})$ | $Q = 4.44\,(\text{Å}^{-1})$ | $Q = 3.14\,(\text{Å}^{-1})$ | $Q = 4.44\,(\text{Å}^{-1})$ |
| PG(002) | 40'-40'-40'-40' | 941 | 455 | 73 | 31 |
| PG(002) | 10'-40'-20'-40' | | | 52 | 18 |
| PG(002) | 10'-10'-10'-10' | 100 | 50 | 23 | 9 |
| Ge(004) | 40'-40'-40'-40' | | | | 20 |
| Ge(004) | 10'-40'-20'-40' | | 15 | | 12 |
| Ge(004) | 10'-10'-10'-10' | | | | 5 |
| Ge(220) | 40'-40'-40'-40' | | | 45 | |
| Ge(220) | 10'-40'-20'-40' | 41 | | 33 | |
| Ge(220) | 10'-10'-10'-10' | | | 10 | |

We note that different sample mosaics can lead to different relative intensities. With a normal (1° mosaic) single crystal sample, the calculated intensity using a Ge(220) or (004) perfect crystal analyzer with beam collimations of 10'-40'-20'-40' is about 2 to 5 times smaller than that using a PG(002) analyzer with 10'-10'-10'-10' collimations. On the other hand, when the sample crystal has a very fine mosaic, e.g., 1', as used in our calculations, the calculated intensities between the perfect Ge analyzer set-up and PG(002) set-up are not very different.

Because of the huge extinction effects present when using a perfect crystal as the analyzer, the analyzer reflectivity can not be accurately estimated. The extinction from the Ge perfect crystal analyzer can greatly reduce the intensity, thus making it impossible to compare intensities between different analyzer set-ups directly. Nevertheless, a comparison between different collimations using the same analyzer reflection can be quite informative.

For example, with Ge (004) reflection as the analyzer, the intensity around $q = 4.44$ Å$^{-1}$ increases by a factor of two going from 10'-10'-10'-10' to 10'-40'-20'-40'. On the other hand, the resolution coarsens with the collimations, but it depends much on the range of $q$ being measured. Therefore, one needs to carefully consider the trade-off between resolution and intensity gain/loss in order to choose the most appropriate collimation settings.

## 4.   Examples

In this section, we provide some examples of using this powerful new technique in our measurements.

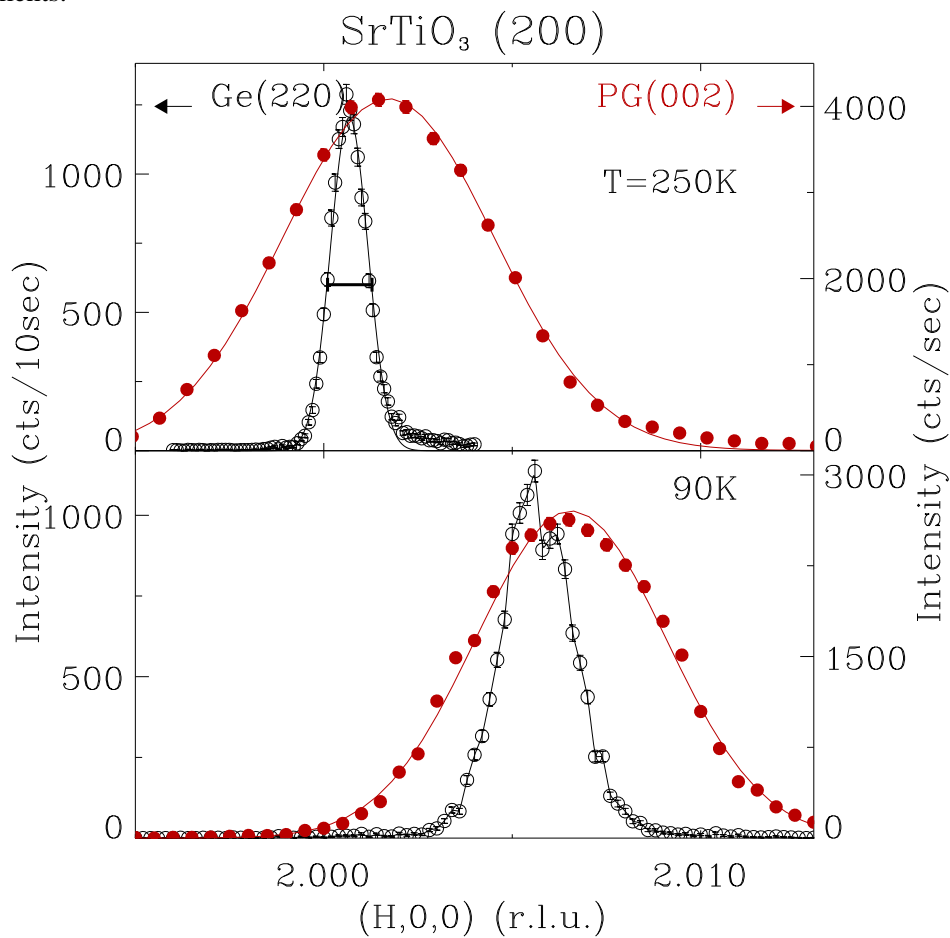

Fig. 8. Longitudinal profiles of the (200) Bragg peaks of SrTiO$_3$ measured at 250 K (top frame) and 90 K (bottom frame). The horizontal bar indicates the $q$-resolution along the (200) direction. The incident neutron energy is $E_i = 14.7$ meV. The collimations are 10'-40'-20'-40' for the Ge(004) set-up and 10'-10'-10'-10' for the PG(002) set-up.

Fig. 8 shows our measurements of the (200) Bragg peak of a SrTiO$_3$ single crystal, using a PG(002) monochromator, tuned to $E_i = 14.7$ meV. Results using a perfect crystal Ge(220) analyzer (beam collimations of 10'-40'-20'-40') and PG(002) analyzer (collimations 10'-10'-10'-10') are compared. The measurements are performed on the BT9 triple axis spectrometer located at the NCNR.

At T=250 K, the system is in a cubic phase and the Bragg profile is a very sharp Gaussian-shaped peak. With the Ge(220) set-up, the peak width $\delta q/Q \sim 6 \times 10^{-4}$, is in good agreement with the calculated resolution width ($\delta q/Q \sim 5.5 \times 10^{-4}$).

Below the structural phase transition $T \sim 110$ K, the system transforms into a tetragonal phase, and a splitting of the (200) Bragg peak occurs. The splitting $c/a \approx 1.0005$ (Hirota *et al.*, 1995) is very small and almost impossible to measure by conventional neutron triple-axis techniques. We measured the (200) peak with the PG(002) analyzer and the best achievable beam collimations of 10'-10'-10'-10', yet no splitting can be detected. With our perfect Ge(220) crystal analyzer set-up, one can clearly see that the original single peak splits into two separate peaks, with $\Delta q \sim 0.001 (r.l.u.)$ ( $\Delta q/Q_{(200)} \sim 5 \times 10^{-4}$)

We have also performed transverse scans at the SrTiO$_3$ (200) Bragg peak, with both the Ge(220) and PG(002) set-ups (Fig. 9). The measurements were done in the cubic phase at T= 250 K. With the Ge(220) analyzer, the scan shows a very sharp peak, with a width $\delta q/Q \sim 6 \times 10^{-4}$, slightly broader than the calculated transverse resolution (horizontal bar in the figure) $\delta q/Q \sim 4.5 \times 10^{-4}$, indicating a small finite sample mosaic. Even so, it is still about three to four times sharper than that measured by the 10'-10'-10'-10' collimated PG(002) set-up.

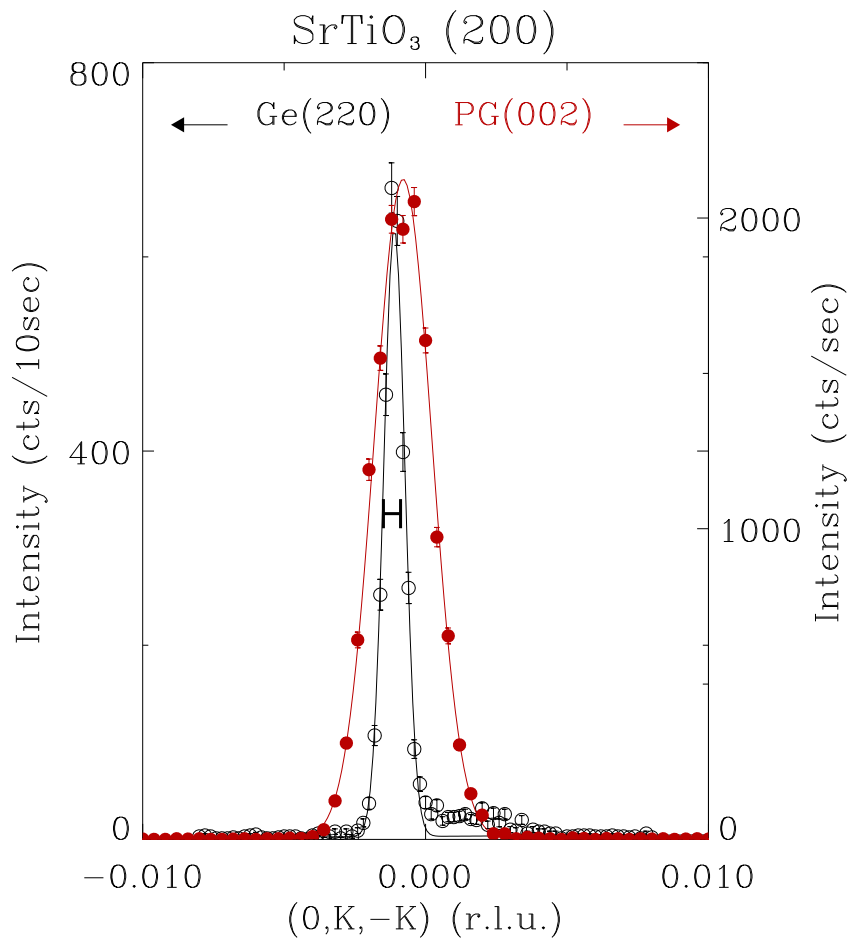

Fig. 9. Transverse profiles of the (200) Bragg peaks of $SrTiO_3$ measured at 250 K. The solid lines are fits to gaussian peaks. The horizontal bar indicates the $q$-resolution along the (220) direction. The incident neutron energy is $E_i = 14.7$ meV. The collimations are 10'-40'-20'-40' for the Ge(220) set-up and 10'-10'-10'-10' for the PG(002) set-up.

During our recent work on the relaxor perovskite $PbXO_3$ systems, we employed this technique using the (004) reflection of a perfect Ge crystal as the analyzer. With larger $\tau$, a better relative Bragg width has been achieved. In Fig. 10, we show longitudinal scans at a (220) Bragg peak of $PbXO_3$ (X=$(Mg_{1/3}Nb_{2/3})_{0.73}Ti_{0.27}$) (Xu $et$ $al.$, 2003). Here $q \approx 4.4$ Å$^{-1}$, very close to $\tau_{Ge(004)} = 4.442$ Å$^{-1}$. For T above the Curie temperature $T_C \approx 375$ K, the system is in cubic phase and we have a very sharp (220) peak, $\delta q/Q \sim 3 \times 10^{-4}$. With the Ge(004) set-up, we can achieve a relative Bragg width almost half of that using the Ge(220) reflection.

For T below $T_C$, the system transforms into a rhombohedral phase, and a splitting of the (220) Bragg peak occurs. The splitting of the two peaks is $\Delta q/Q_{(220)} \sim 2.5 \times 10^{-3}$, and the width of the two split peaks is $\delta q/Q_{(220)} \sim 2.2 \times 10^{-3}$. With our perfect Ge(004) crystal analyzer set-up, not only the splitting, but also the broadening of the two peaks, which is a result of the internal strain, can be accurately measured.

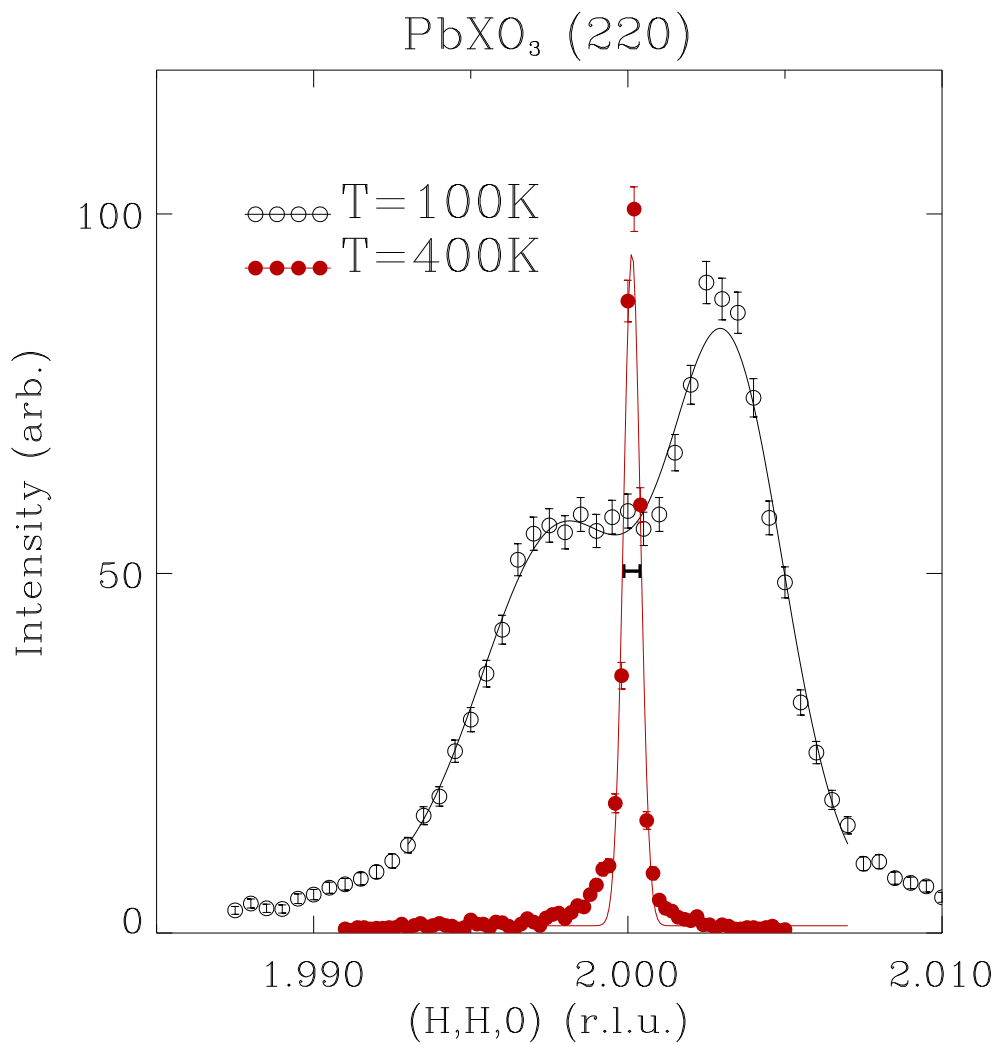

Fig. 10. Profiles of the (220) Bragg peaks for PbXO$_3$ (X=(Mg$_{1/3}$Nb$_{2/3}$)$_{0.73}$Ti$_{0.27}$) measured with a Ge(004) perfect crystal analyzer, are shown at 100 K (open circles) and 500 K (solid circles). The beam collimations are 10'-40'-20'-40 and E$_i$ = 14.7 meV. The solid lines are fits to Gaussians. The horizontal bar indicates the $q$ resolution along the (220) direction.

# 5. Summary

Based on the previous calculations and experimental examples, we show that a huge improvement in the relative Bragg width can be achieved when the analyzer-side focusing condition is satisfied by matching the measurement wave-vector to a perfect crystal analyzer reflection. With this new technique, one can still enjoy the flexibility of a triple-axis-spectrometer, while improving the moderate relative $q$-resolution of conventional TAS by one to two orders of magnitude.

Due to the primary extinction from the perfect analyzer crystal, the Bragg intensity using this new technique is usually a few orders of magnitude smaller than a conventional PG(002) analyzer set-up (see Fig. 2). For elastic measurements, this is usually not the problem if using a reasonable sized single crystal. By using different collimations, the intensity/resolution trade-off can be tuned to best fit the desired measurements.

We would like to thank C. Stock, S. Wakimoto, and Z. Zhong for stimulating discussions. Financial support from the U.S. Department of Energy under contract No. DE-AC02-98CH10886 is also gratefully acknowledged.

---

## Synopsis

A new technique which brings a substantial increase of the wave-vector $q$-resolution of triple-axis-spectrometers is introduced.